\documentclass[aps,prb,preprint,groupedaddress]{revtex4-1}
\usepackage{graphicx}
\usepackage{epsfig}
\usepackage{subfigure}

\begin{document}
\title{Structural and phonon properties 
of bundled single- and double-wall carbon nanotubes under pressure}

\author{A. L. Aguiar$^{1,2}$, Rodrigo B. Capaz$^3$, A. G. Souza Filho,$^1$ \\
A. San-Miguel$^2$}

\affiliation{$^1$Departamento de F\'{\i}sica, Universidade Federal
do Cear\'a, P.O. Box 6030, 60455-900 Fortaleza, Cear\'a, Brazil}
\affiliation{$^2$Laboratoire de Physique de la Mati\`{e}re Condens\'ee et
Nanostructures, Universit\'e Claude Bernard Lyon-1 et CNRS, 69622 Villeurbanne Cedex, France}
\affiliation{$^3$Instituo de F\'{\i}sica, Universidade Federal
do Rio de Janeiro, P.O.Box 68528, 22941-972, Rio de Janeiro, RJ, Brazil}

\date{\today}

\begin{abstract}
In this work, we report a theoretical coupled study of the structural and
phonons properties of bundled single- and double-walled carbon nanotubes (DWNTs),
under hydrostatic compression.  
Our results confirm drastic changes in volume of SWNTs in high-pressure regime 
as assigned by a phase transition from circular to collapsed phase which are strictly dependent on the tube diameter. 
For the DWNTs, those results show first 
a transformation to a polygonized shape
of the outer tube and subsequently the simultaneous collapse of the outter and inner tube, at the onset of the inner tube
ovalization. Before the DWNT collapse, phonon calculations reproduce the experimentally observed screening effect
on the inner tube pressure induced blue shift both for RBM and tangential G$_z$ modes . 
Furthermore, the collapse of CNTs bundles induces a sudden redshift of tangential component in agreement with
experimental studies. The G$_z$ band analysis of the SWNT collapsed tubes shows that the flattened regions of the tubes are
at the origin of their G-band signal. This explains the observed graphite type pressure evolution of the G band in the collapsed phase
and provides in addition a mean for the identification of collapsed tubes.  

\end{abstract}

\pacs{3333333}

\maketitle

\section{Introduction}

Carbon nanotubes are fundamental materials for developing nanoeletromechanical systems because 
they present a unique combination of outstading electronic and mechanical properties.
Cross-sectional deformations of single-wall carbon nanotubes (SWNTs) under pressure have 
been extensively studied by means of experimental tools \cite{TangPRL00,peters00,karmakar03,merlenPRB05,yao08} and 
theoretical models. \cite{CharlierPRB96,YildirimPRB00,sluiterPRB02,reichPRB02,CapazPSS04,tangney05,ChoiPSS07} 
On the other hand, the 
behavior of double-wall carbon nanotubes (DWNTs) under pressure has been considerably less 
studied \cite{arvanitidis05,puechPRB06,puechPRB08,kawasaki08}. DWNTs may be better candidate 
materials than SWNTs for the engineering of nanotube-based composite materials because of their 
geometry, in which the outer tube ensures the chemical coupling with the matrix and the inner 
tube acts as mechanical support for the whole system\cite{aguiar11}.

Some theoretical studies have been performed to study the pressure dependence of the 
cross-sectional shape of SWNT, depending on the chirality and diameter. 
By starting from an almost perfect circular cross section, the cross-sectional shape of the nanotube
becomes oval or polygonized as the pressure is increased, evolving later to a collapsed state with a peanut shape.
\cite{tangney05,ImtaniPRB07,ImtaniCMS08,ImtaniCMS09,yangPRB07} 
Based on elasticity theory, Sluiter et al. have proposed a diameter-dependent phase diagram for 
the cross-sectional shape of SWNTs under pressure.\cite{SluiterPRB04}
Some authors reported that the radial deformation of SWNTs for diameters smaller than 2.5nm is reversible from the collapsed state while 
the deformation of larger diameter tubes could be irreversible and the collapsed state is metastable or even
absolutely stable without pressure application. \cite{tangney05,SluiterPRB04} 
Many experimental studies have suggested that phase transition 
in nanotubes could be dependent on their metallic character or on the 
surrounding chemical environment used for transmitting the pressure. \cite{christofilosDREL06,merlenPSS06,proctorPSS07,merlenPRB05,AbouelsayedJPCC10} 
However, theoretical calculations suggest that phase transitions of SWNTs under pressure 
is mainly dependent on the cube inverted diameter (p$_c$ $\sim$ d$_t^{-3}$) of the tubes and not on the chirality. \cite{tangney05,elliot04}
Even if there is a huge dispersion of results concerning 
the pressure transition values from theoretical models, there is an overall agreement
between different calculations on the existence of two phase transitions (circular-oval and oval-peanut) 
and that the critical pressures for those transitions are diameter dependent.

The pressure evolution of DWNTs has been much less studied and, in opposition to SWNTs, a detailed
understanding of their cross-sectional evolution is still under discussion due their complexity regarding the role of inner
and outer tubes. Ye et al. has
suggested that the critical collapse pressure of DWNT is essentially determined by inner tube stability and
so the collapse pressure of DWNT is close to what is expected for inner tube \cite{YePRB2005}. Other authors
suggested that collapse pressure of DWNTs is completely different from which is expected for the correspondent SWNT
when are considerate separately and, the pressure value still depends on 1/d$_t^{*3}$ scale law but with a suitable choice of an 
average diameter d$_t^*$ \cite{yangAPL06,gadagkar06}. In this paper we report a study of the 
vibrational properties of carbon nanotubes bundle under pressure, a subject that has not been theoretically well explored in the literature
up to now.

This paper is organized as follows. First, we will describe in section II the methodology used to model nanotube bundled structure. 
In section III.A, we calculated the structural evolution of
SWNTs and DWNTs bundles under pressure. We focused in structural stability of bundle structure by calculating the critical pressure
for collapsing and we compare the behavior of DWNT bundle with its corresponding SWNT. In section III.B, we 
explored the vibrational properties of
SWNT and DWNTs bundle under pressure by calculating RBM and tangential phonons modes before and after the nanotube collapse.
We close our paper with conclusions in section IV.

\section{Methodology}

In order to access mechanical and vibrational properties of carbon nanotubes under pressure 
we initially perform zero-temperature structural minimizations of SWNTs and DWNTs bundles. 
The carbon-carbon bonding within each CNTs is modeled by a reactive empirical bond order 
(REBO) potential proposed by Brenner \cite{brenner02,brenner90}. Pairwise Lennard-Jones potentials are added to model 
the non-bonding van der Waals terms ($\epsilon/k_b$=44K, $\sigma=3.39$ \AA),
which are essential to describe intertube interactions \cite{tangney05}. We studied zigzag and armchair C
bundles of CNTs in triangular arrangements within 
orthogonal unit cells containing two SWNTs or DWNTs each, with lateral lattice constants $a_x$ and $a_y=\sqrt{3}a_x$ 
(the $a_y/a_x$ ratio is kept fixed). The lattice constant along the z axis, ($a_z$), is chosen to contain 5 (8) 
unit cells of the zigzag (armchair) tubes.  

We perform a sequence of small and controllable steps of unit cell volume reduction for each
fixed nanotube phase studied (circular, polygonized, oval, peanut,etc.). 
Each step is of the order of $\Delta V/V_0$ = -0.01$\%$ , where $V_0$ is the ambient pressure volume
for each nanotube system in circular phase. For each fixed volume $V$, we search for the atomic positions and 
lattice constants $a_x$ and $a_z$ that minimize the 
internal energy $U(V)$. The pressure is obtained by $p=-\Delta U/\Delta V$ 
and the enthalpy is $H = U + pV$.

Phonon frequencies and eigenvectors are directly obtained by the diagonalization of the force constant 
matrix. The matrix elements
are calculated by using finite difference techniques. We focus our vibrational analysis on the 
radial breathing mode (RBM) and longitudinal G-band (denoted G$_z$), whose displacements are along the tube axis. 
Therefore, RBM and G$_z$ projected density of states (PDOS) are constructed by projecting 
the phonon eigenvectors onto the corresponding radial and axial (with opposite phases 
for atoms in different sub-lattices) displacement fields, respectively.
  
\section{SWNTs and DWNTs under pressure: Structural and Vibrational Properties}

\subsection{Nanotube Collapse}

\begin{figure}[ht]
\includegraphics[scale=0.15]{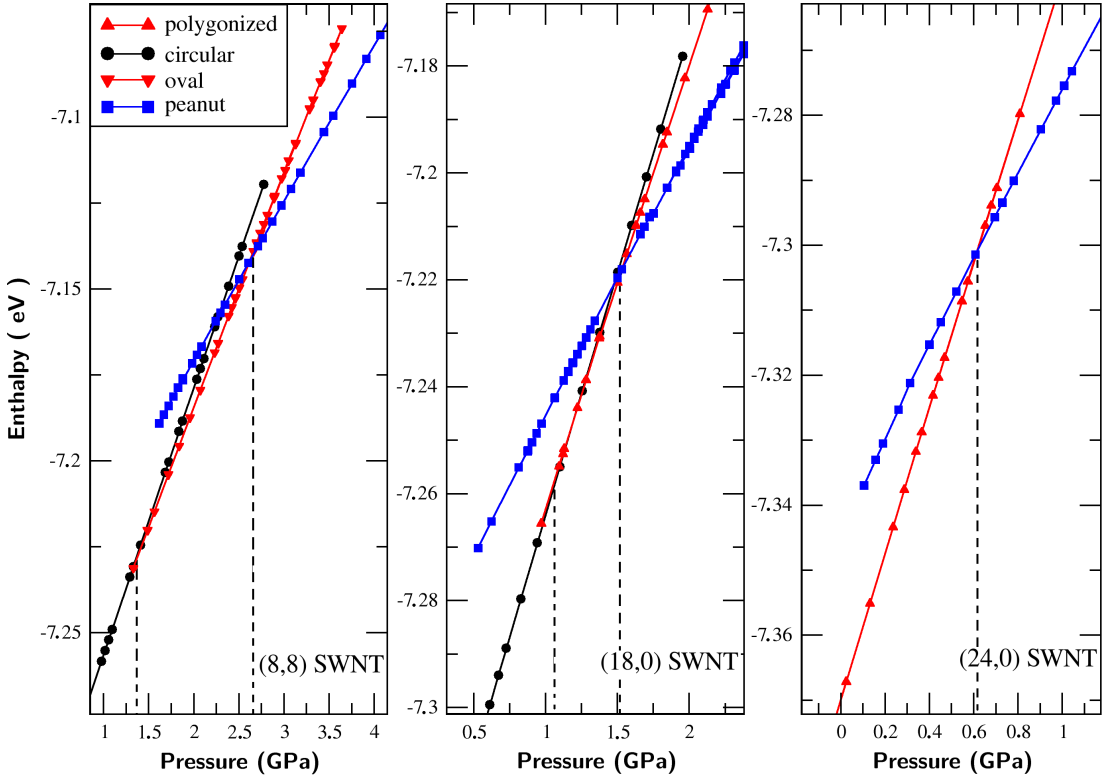}
\caption{\small{Enthalpy vs pressure curvers calculated for (a) (8,8), (b) (18,0) and (c) (24,0) bundled SWNTs.
The pressure evolution for (10,0) SWNT is similar to (8,8) SWNT showed in (a) and the circular-oval and oval-peanut transitions
is found close to 1.55 and 9.6 GPa}}
\label{Fig1-swnt}
\end{figure}

We start by studying the low-pressure behavior of SWNT bundles, considering 
four basic cross-sectional shapes namely: circular, oval, polygonal 
(hexagonal) and peanut-like. Tubes of different diameters exhibit a different sequence 
of cross-sectional shapes as a function of pressure \cite{tangney05}. In particular, 
the polygonal or hexagonal phase is typical of bundled tubes and it is often obtained in calculations 
for large diameter nanotubes since plane-parallel facing between adjacent tubes tends 
to decrease the van der Walls interaction energy  \cite{Charlier96,Liu05,RuPRB00}. 
Fig \ref{Fig1-swnt}a shows the enthalpy as a  function of pressure calculated for circular, 
oval and polygonized phases for bundles of (8,8), (18,0) and (24,0)  SWNTs. The crossing of two curves 
in this plot by following the lowest enthalpy for a given pressure indicates a transition 
between two different cross-sectional shapes. The derivative discontinuities in the $H(p)$ 
plots are associated with the volume variation at the phase transition. This can be seen more 
clearly when $p-V$ plots are constructed (not shown here) where each 
transition is marked by a discontinuous change in volume.
For the (8,8) SWNT bundle, the transition pressure from circular to oval (so called p$_1$) occurs close to 1.5 GPa 
and from oval to peanut (so called p$_2$) around 2.6 GPa. For the (18,0) SWNT bundle, 
we found phase transitions at 1.2 GPa (circular to polygonal, p$_1^\prime$) and 1.5 GPa (polygonal to peanut, p$_2^\prime$). 
Finally, for the (24,0) SWNT bundle the polygonal to peanut transition (p$_2^\prime$) is found at 0.6 GPa. 
We see that the small diameter 
(8,8) SWNT bundle undergoes a circular$\rightarrow$oval$\rightarrow$peanut phase transition sequence as the pressure increases, 
whereas for the intermediate diameter (18,0) SWNT the sequence is circular$\rightarrow$polygonal$\rightarrow$peanut and 
for the large-diameter (24,0) we obtain simply a polygonal-peanut transition, since the tubes are 
already polygonized even at zero pressure.

As expected, the critical pressures are strongly diameter-dependent \cite{benedict98,tangney05,sun04,elliot04,ImtaniPRB07,ImtaniCMS09}. 
Actually, many authors have pointed out that
the critical pressure ($p_1$) for circular-oval transition scales with d$_t^{-3}$, where d$_t$ is tube diameter.\cite{tangney05} 
Some authors also found that the critical pressure for oval-peanut transition ($p_2$) also scales with d$_t^{-3}$. \cite{elliot04,wuPRB04}. 
Our results for bundled SWNTs shown in Fig \ref{Fig1-swnt} point out that the transition to the peanut geometry 
($p_2$ and $p_2^{'}$) follows approximately 
the same law $p_2=C/d_t^3$ with $C$=4.4nm$^3$.GPa. This scaling law was observed for all 
studied zigzag and armchair bundled tubes. However, the pressure transition values $p_1$ e $p_1^{'}$ for circular to intermediate phase 
(oval and polygonal, respectively) show a slightly dependence on the inverted tube diameter but
they do not follow the d$_t^{-3}$ scaling law.

\begin{figure}
\centering
\includegraphics[scale=0.6]{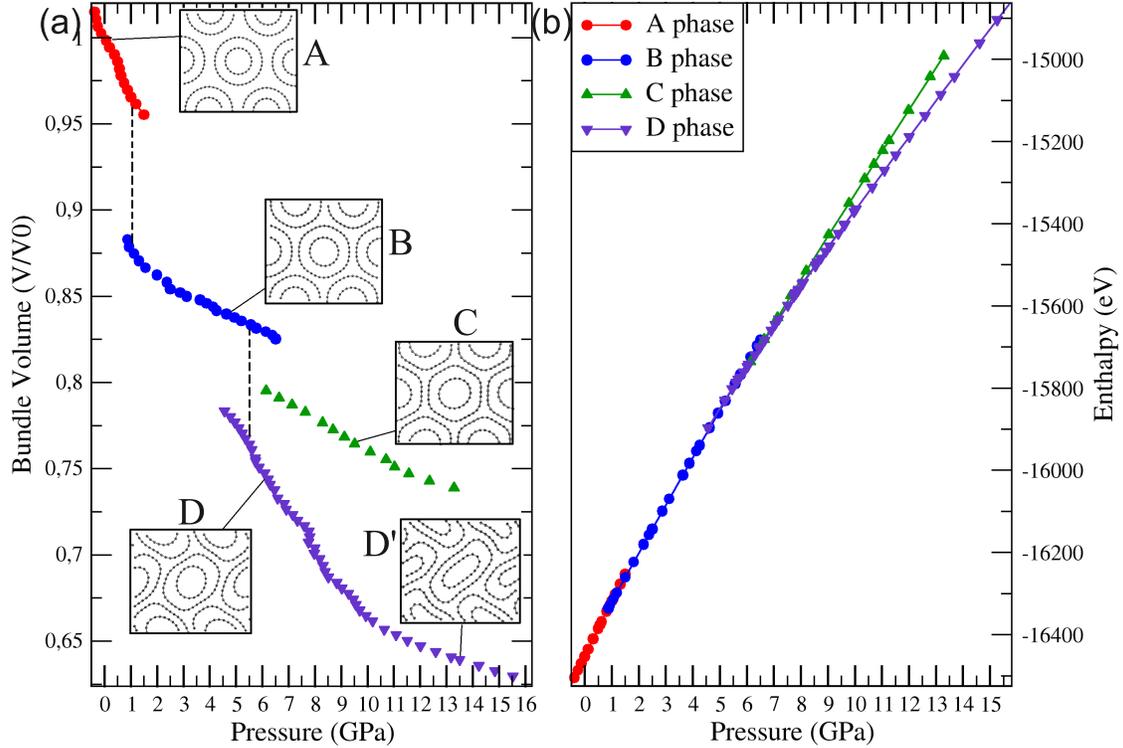}
\caption{\small{p-V curves (a) and enthalpy functions (b) for the 
pressure-induced sequence of phase transitions in the (10,0)@(18,0) DWNT. 
Different cross-sectional shapes correspond to different colors in the plots and are shown in the insets.}}\label{Fig4-dwnt}
\end{figure}

We also have modeled several DWNT bundles under pressure and choose the (10,0)@(18,0) DWNT 
(This notation means that the (10,0) tube is inside the (18,0) tube)
in order to compare its 
structural stability with those of their constituent SWNTs. We find a very similar behavior for the other DWNTs
such (12,0)@(20,0) and (11,0)@(19,0). 
In Fig. \ref{Fig4-dwnt}a and \ref{Fig4-dwnt}b we show respectively the $p-V$ and enthalpy curves for the (10,0)@(18,0) DWNT bundle.  
Upon compressing the original circular structure of a DWNT we find four distinct structures
 as shown in Fig. \ref{Fig4-dwnt}a. The $A$ configuration is the
non-deformed structure with a circular shape for both inner and outer tube. The $B$ configuration is 
characterized by the polygonization of the outer tube while the inner tube maintain its circular cross section.
The $C$ configuration corresponds to a polygonal outer tube and oval inner tube. Finally, the $D$ configuration 
stands for the case where the outer and inner tubes display oval/peanut cross-sectional shapes.

The polygonization of outer tube ($A$-$B$ phase transition) in (10,0)@(18,0) DWNT bundle is observed at 0.81 GPa which is 
lower than the value (1.2 GPa) what was calculated for the same transition in the (18,0) SWNT bundle. 
This result suggests that the presence of the inner tube enhances the outer-outer 
tube interactions within the bundle, thus reducing
the critical pressure value for polygonization. Upon increasing the pressure, a $B$-$C$ transformation 
occurs at 6.2 $\pm$ 0.2 GPa, since the inner tube ovalizes and the outer tube reaches a higher 
degree of polygonization. We are able to increase the pressure in the $C$ phase up to 13 GPa without inducing any
strong deformation. However, from the enthalpy analysis 
we conclude that the $D$ phase is more stable than $C$ phase for the whole range of investigated pressures.
Thus the $C$ configuration is actually metastable. By further increasing the pressure in the $D$ phase, it reaches 
the same enthalpy as the $B$ phase around 5.7 $\pm$ 0.2 GPa. The $D$ phase is also metastable even 
at lower pressures, down to 5.5 GPa. In summary, the actual sequence of phase transitions for the (10,0)@(18,0) 
DWNT bundle is $A$ $\rightarrow$ $B$ at 0.81 GPa and $B$ $\rightarrow$ $D$ at 5.7 $\pm$ 0.2 GPa. This value contrasts with the 1.4 GPa
calculated collapse pressure of the (18,0) SWNT outer tube, cleary poiting out to the structural support
given by the inner tube, as already proposed by some authors \cite{yangAPL06,gadagkar06}.

We conclude that the $C$ configuration (polygonized outer tube and ovalized inner tube) is 
unstable when we compare with the both collapsed tubes which let us propose that the ovalization of the inner tube is
the reason for the overall collapsing of DWNT. For isolated DWNTs, Ye et al \cite{YePRB2005} suggest that the DWNT transition could be 
essentially determined by the inner tube transition.
They also found that collapse of DWNTs follow almost the same d$_t^{-3}$ law when the inner tube is used, which means that p$_c$ for
DWNTs is the same as expected for SWNT inner tube \cite{YePRB2005}. In such way, our results agrees in the 
fact that the inner tube ovalization determines the critical
pressure for collapsing of DWNTs. However, there is a clear pressure screening effect from the outer tube in the behavior 
of the inner tube because the critical pressure for the expected circular-oval 
transition for inner tube was increased to 5.7 $\pm$ 0.2 GPa when it was encapsulated in the DWNT. This value is considerably higher 
than that observed for the (10,0) SWNT bundle, which was 1.55 GPa. 
Second, the full collapsing of the inner tube (oval-peanut transition) is continuous in the DWNT case, 
differently from the SWNT case, in which there is a discontinuous change in the volume (around 9.6 GPa).
For the DWNT case, this transition is marked by a change in compressibility (which is proportional to the slope of the $p-V$ plot) 
at around 10 GPa, clearly seen from Fig.\ref{Fig4-dwnt}. We found a similar behavior for (12,0)@(20,0) DWNT bundle, 
where the transition $B$ $\rightarrow$ $D$ is predicted to occur at 5.5 $\pm$ 0.1 GPa
and change in bulk modulus takes places around 5.9 GPa close to that value expected for (12,0) SWNT (5.4 GPa).


\subsection{Phonon Density of States evolution under pressure}
\begin{figure}[ht]
\centering
\includegraphics[scale=0.19]{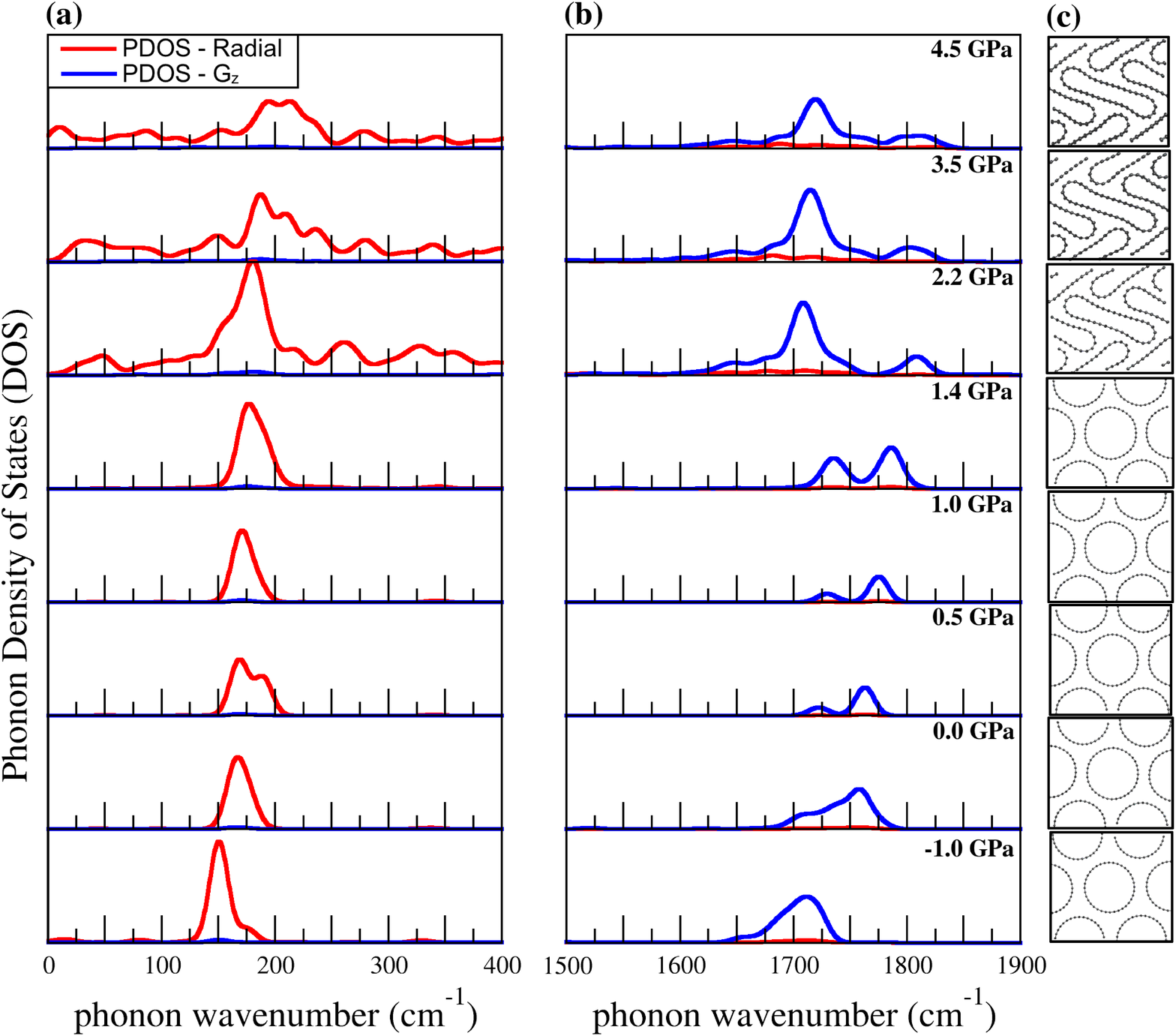}
\caption{\small{Phonon density of states (DOS) projected in RBM (a) and tangential (b) probing vectors for (18,0) SWNT.
Some snapshots of the bundle structure evolution can be followed in (c).}}\label{Fig18-ph}
\end{figure}

Fig \ref{Fig18-ph} shows the phonon projected density of states (PDOS) calculated for (18,0) SWNT bundles. 
The PDOS for RBM and G$_z$ modes are shown in red (a) and blue (b), respectively. From the figure, we can follow 
the RBM and  G$_z$ evolutions as pressure is increased, 
and the corresponding snapshots of the bundle structural evolution are shown in the right panels. First, we observe
the RBM mode of (18,0) SWNT in the low frequency region centered at around 150 cm$^{-1}$ in the circular phase (0.0 GPa), which gradually
shifts to higher frequencies as the pressure is increased. The pressure coefficient of this mode is 7.0 $\pm$ 2.5 cm$^{-1}$/GPa, in good 
agreement with experiments \cite{caillier08,merlenPSS06,peters00}.
After collapse, the contributions to the radial mode spread in the low frequency region, which is equivalent to say that the RBM is 
not a single, well-defined mode any longer in good correspondance with the experimental difficulty to observe RBM after the 
some suggested pressure phase transitions. \cite{peters00,karmakar03,sandler03,elliot04,freire07,Proctor06}

The G$_z$ band behavior is shown in the right panel. We clearly observe the splitting of this mode when 
polygonization takes place.  
After collapse, we clearly see a sudden jump to lower frequencies and intensity enhancement of the 
lower-frequency contribution, while the higher-frequency one continues to upshift. 
We studied several SWNTs bundles and similar results were observed. 
For (10,0) SWNT bundle the spread of
radial contribution were observed in circular-oval transition i.e. before the collapsing of tube. The split of G$_z$
component due polygonization is also observed for (24,0) SWNT bundle and, evidently, not observed for (10,0) SWNT. Furthermore,
the sudden red-shift of G$_z$ component after collapsing is also observed for (24,0) SWNT.

\begin{figure}[ht]
\centering
\includegraphics[scale=0.50]{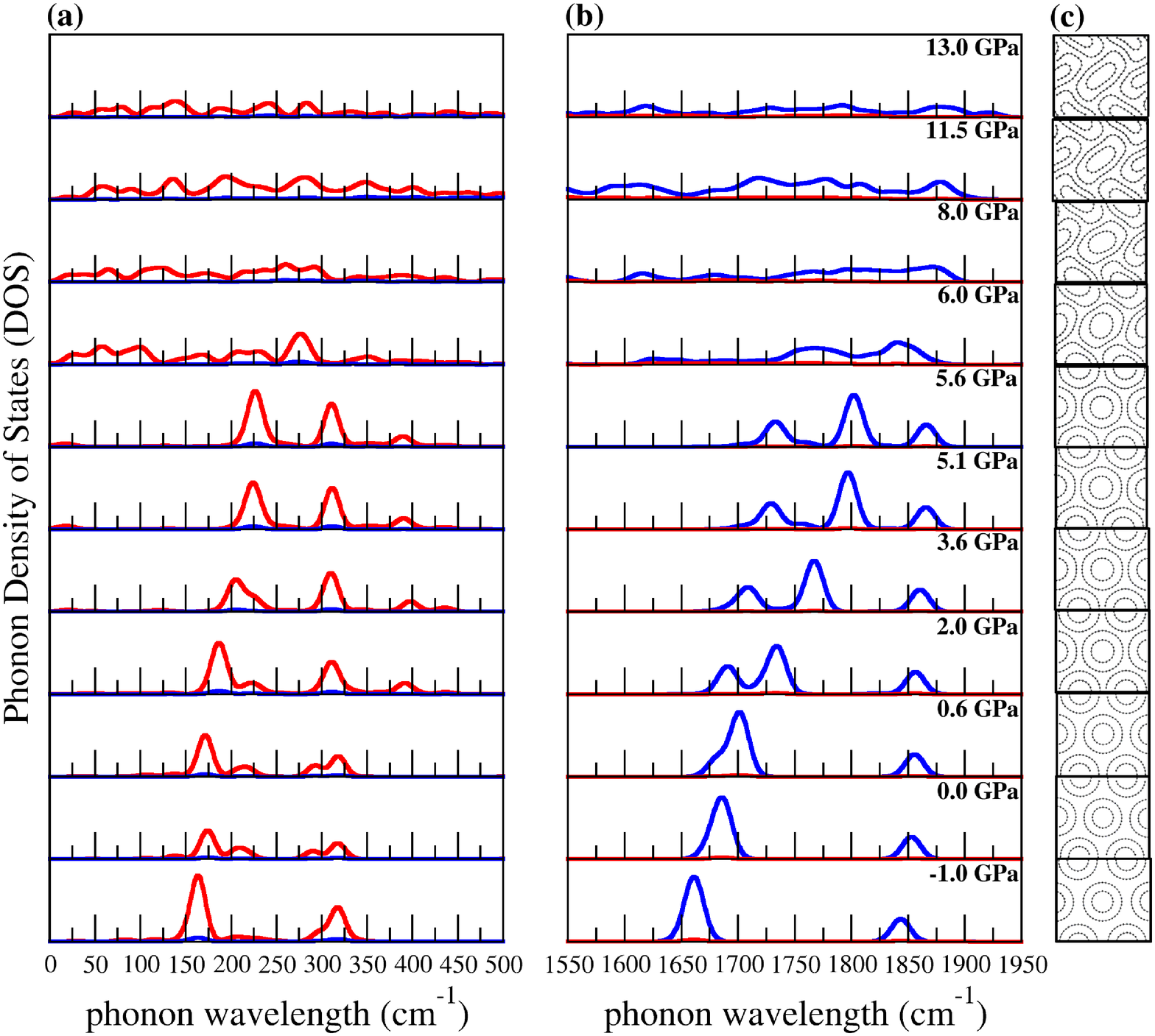}
\caption{\small{Phonon density of states (DOS) projected in RBM (a) and tangential (b) probing vectors for (10,0)@(18,0) DWNT.
Some snapshots of bundle evolution can be followed in (c).}}\label{Fig1810-ph}
\end{figure}

The phonon evolution with pressure for (10,0)@(18,0) DWNT bundle can be observed in Fig \ref{Fig1810-ph}. It is clear from left panels
the presence in low-requency region of two distinct peaks around 150 cm$^{-1}$ and 325 cm$^{-1}$ corresponding to RBM modes of outer and
inner tubes, respectively. The peaks are slightly shifted from the corresponding SWNTs in the circular conformation, 
as expected from the intertube coupling \cite{PopovPRB02}. It is interesting to note the evolution 
of both RBM peaks with pressure, where we clearly observe that the RBM frequency of the outer tube shifts much faster than the RBM frequency
of the inner tube before collapse. 
This is a clear evidence of a screening effect on the inner tube by outer tube. Pressure screening effects on RBM and G band 
in DWNTs have been observed in several Raman experiments \cite{christofilos07,aguiar11,puech04,PuechPSS04,arvanitidis05}. Our 
calculations of RBM and G$_z$ confirm that this pressure screening effect occurs well before any structural collapse.
In the (b) panels of Fig. \ref{Fig1810-ph}, we can follow the
G$_z$ component for outer and inner tube. It should be noted here that inner (10,0) tube G$_z$ component is well overestimated being at about
1850 cm$^{-1}$ compared with the outer (18,0) one which is located at around 1650cm$^{-1}$. However, qualitative analysis could be still performed.
It is interesting to note that pressure screening effects are also observed in tangential components as we can see that the outer tube
G$_z$ components shift faster than the inner ones. As observed for SWNTs, the transition of the outer tube to a 
polygonized phase at 0.85 GPa is clearly marked by a splitting of the G$_z$ band. 
After colapse, the PDOS spreads over a large frequency range for radial and tangential contributions 
but it is possible to see that the lowest frequency 
components of G$_z$ are sudden shifted to lower frequencies, as in the case of SWNTs bundles. 

\begin{figure}[ht!]
\centering
\includegraphics[scale=0.65]{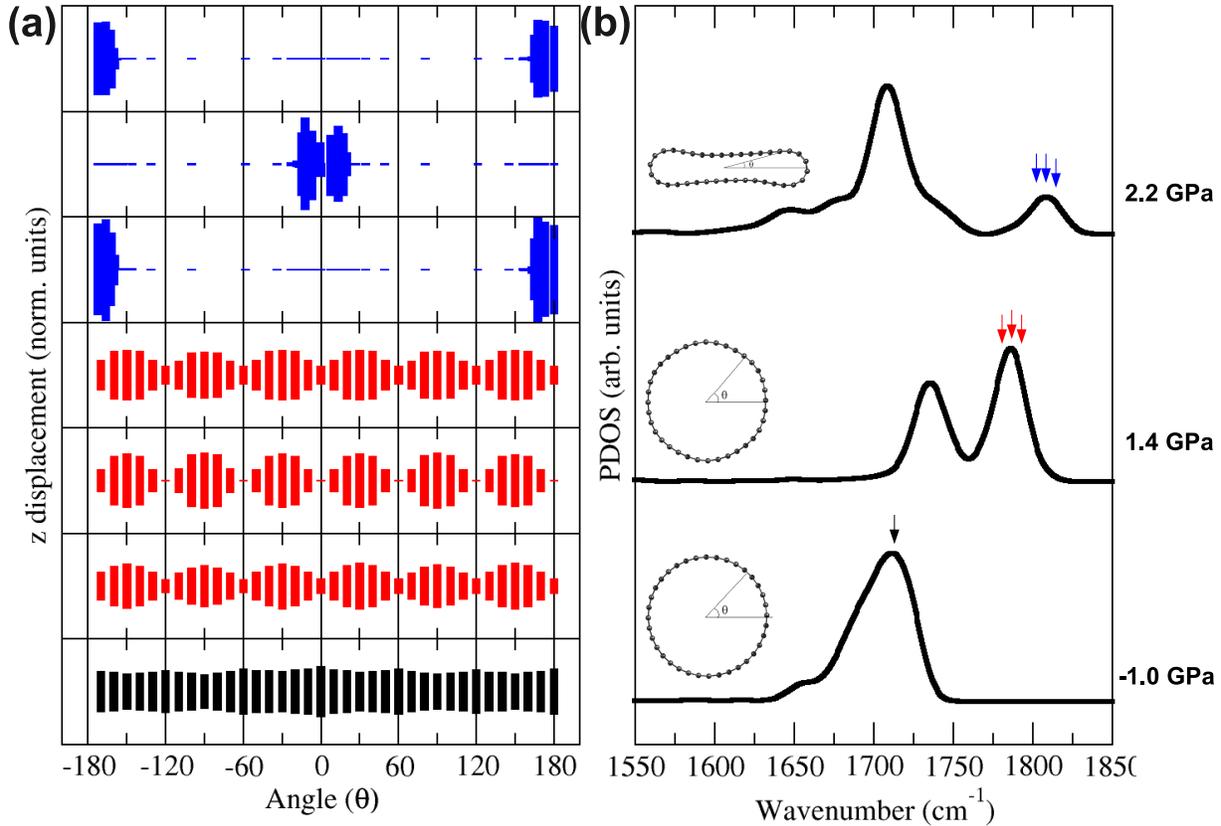}
\caption{\small{(a) Amplitude of z-displacements of (18,0) SWNT bundle as a function of coordinate angle 
before the collapse (-1.0 GPa and 1.4 GPa) and after collapse (2.2 GPa). 
From bottom to top: Eigenvectors for $A_{1g}$ mode centered at 1721.4 cm$^{-1}$ (black) at -1.0 GPa;
for modes centered at 1771.3, 1780.7 and 1786.1 cm$^{-1}$ (red) at 1.4 GPa, and for modes centered at 
1800.6, 1803.2 and 1815.4 cm$^{-1}$ (blue) at 2.2 GPa.
(b) v-PDOS projected in z direction for (18,0) SWNT at -1.0, 1.4 and 2.2 GPa. Arrows mark the center of modes 
whose eigenvectors are plotted in (a).}}\label{rAz}
\end{figure}

In Fig. \ref{rAz}, we plot the z-displacement i.e., along the tube axis direction, of some eigenvectors as a function
of the coordinate angle $\theta$ defined from the center of (18,0) tube. First, we have identified
the $A_{1g}$ mode centered around 1721.4 cm$^{-1}$ for the structure at -1.0 GPa  (Fig. \ref{rAz}a).
This is the mode which mainly contributes to v-PDOS as we can see in Fig. \ref{rAz}b.
As the pressure is increased up to 1.4 GPa, the (18,0) cross section is poligonized as
we observe a split of G$_z$ PDOS contribution. Analysis of the phonon displacements (Fig. \ref{rAz}a) 
shows that the higher frequency component 
is mainly due to modes which are localized on the high-curvature regions (vertices of the hexagons).
Modes centered at 1771.3, 1780.7 and 1786.1 cm$^{-1}$ have the same symmetry and
the their maximums of displacement are localized for $\theta$ equal to $\pm$30$^\circ$, $\pm$90$^\circ$ and $\pm$150$^\circ$
which are the vertices of polygonized shape. After the collapse of (18,0) SWNT bundle,
we have calculated phonon eigenvectors for collapsed structure at 2.2 GPa.
In Fig. \ref{rAz}a, we show modes centered at 
1800.6, 1803.2 and 1815.4 cm$^{-1}$ which mainly contribute to higher-frequency peak in PDOS.
Then, we clearly see that high-frequency contribution in PDOS at 2.2 GPa
come from modes that are localized in the high-curved regions of the peanut-shaped tube. ($\theta$ $\sim$ $\pm$180$^\circ$
and $\theta$ $\sim$ $\pm$0$^\circ$). Consequently, the
low-frequency and more intense peak of PDOS 
is due to vibration modes which are localized in the flat regions of the peanut-shaped structure. 

\begin{figure}[ht]
\centering
\includegraphics[scale=1.10]{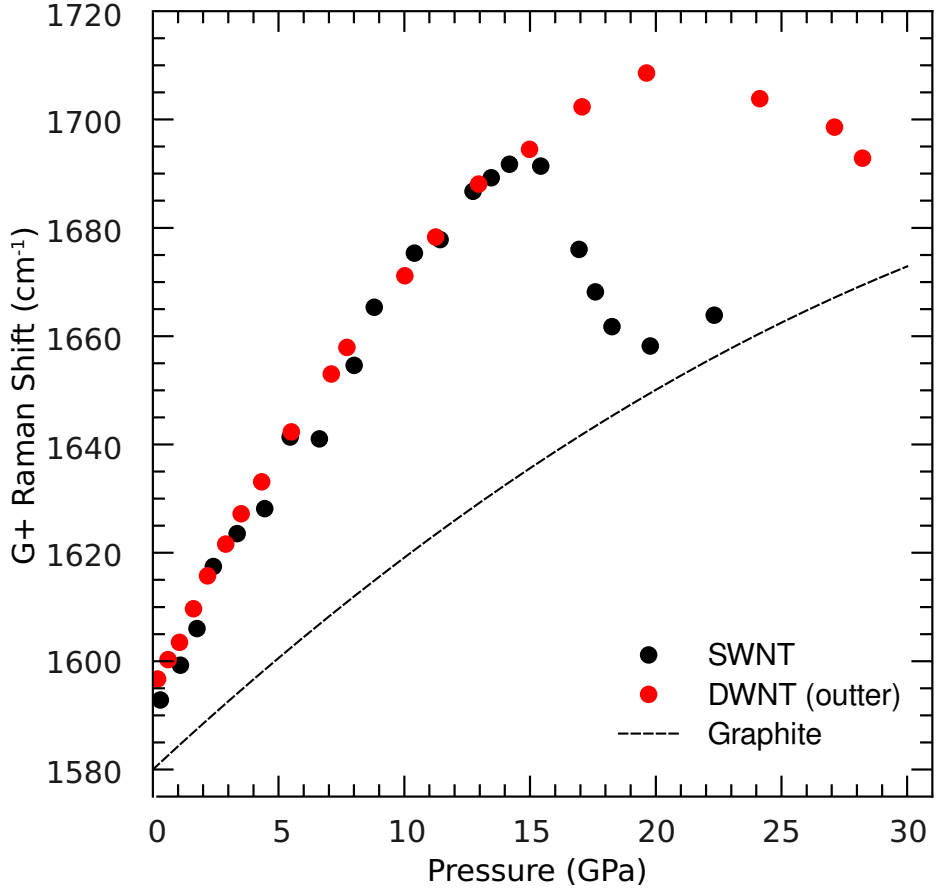}
\caption{\small{Experimentaly pressure Raman shift evolution of G$^+$ component 
for SWNTs, DWNT and graphite. Adapted from Ref. \cite{caillier08,aguiar11,Hanfland89,nicolle11}}}\label{figexp}
\end{figure}

From an experimental point of view, it has been recently proposed that a saturation or even a 
negative pressure slope of Raman G$^+$ component for SWNT and DWNT could be associated to the collapse of the nanotubes \cite{caillier08,yao08,aguiar11}. 
Futhermore, after the collapse this band follows the graphite pressure evolution as observed in Fig. \ref{figexp}.
with a smaller pressure coefficient \cite{aguiar11,Hanfland89}. This was also observed in PDOS of SWNT bundle calculations (cf. Fig \ref{Fig18-ph}b).
Our calculations confirm this hypothesis and identifies the atomistic origin of the low-frequency bands. 
Collapsing of the nanotube cross-section leads to
flattened regions where the reduced stress in C-C tangential bonds reduce the $G_z$ frequency. 
However, as we observe in our calculations, a small high-frequency component of G$_z$ component, 
arising from the high-curvature regions, remains after the collapse.
Since the flattened portion of the tubes is much higher than curved portion, we expect that the low-frequency shift of the G band is 
dominant in the experiments. The experimentaly red-shift observed of the G-band is then explained from our calculations
as related to the dominance of the flat region of the collapse phase in the Raman signal. This is in addition perfectly
coherent with the fact in the collapsed state the pressure evolution of the G-band clearly matches (see Fig. \ref{figexp})
the one of graphite or graphene under triaxial compression \cite{Hanfland89,nicolle11}.

\section{Conclusions}

We studied high-pressure structural modifications and phonon-mode shifts of single and double wall carbon nanotubes bundles using 
zero-temperature enthalpy minimization with classical interatomic potentials. We confirmed that the structural transformation
of SWNTs in bundles from circular to collapsed cross-section under hydrostatic pressure is strongly diameter-dependent. 
It is also evident that the transition from polygonized to collapsed
cross section for large diameter tubes has a first-order character with a large 
bundle volume discontinuity whereas for small diameters the transition is continuous, with 
intermediate oval or racetrack cross sections. For DWNT bundles, screening pressure effects were observed on the inner tube, which keeps 
its circular cross section for higher pressures than would be expected from the SWNTs 
behavior. Furthermore, the inner tube acts as structural support to outer tube.
Phonons calculation also reveals screening effects in radial and tangential 
component as the pressure increases. Polygonization of (18,0) SWNT bundle is characterized by
a split of G$_z$ PDOS contribution which is mainly due to localized modes on high-curved region of polygonal
shape. After the collapse transition, the tangential modes associated 
with the flat regions of the tubes are suddenly shifted to low frequencies and high-curved contribution
of peanut shape is less pronounced but still observed in G$_z$ PDOS. The experimentaly observation of the G-band
red-shift at nanotube collapse let us conclude on the dominance on the Raman signal of collapsed tubes
of the flat regions of collapsed tubes. Moreover, this provides a coherent explanation of the lower pressure slope
($\sim$ 4cm$^{-1}$/GPa) of the G-band observed after nanotubes collapse. Consequently, the pressure slope could
be a useful means for the identification of the geometry shape state of SWNT and DWNTs. 

\section*{Acknowledgements}
The authors acknowledge CAPES-COFECUB collaboration program (grant 608) grant for the partial support of this research.

\bibliographystyle{unsrt}

\bibliography{minebib3}

\begin{thebibliography}{10}

\bibitem{TangPRL00}
Jie Tang, Lu-Chang Qin, Taizo Sasaki, Masako Yudasaka, Akiyuki Matsushita, and
  Sumio Iijima.
\newblock Compressibility and polygonization of single-walled carbon nanotubes
  under hydrostatic pressure.
\newblock {\em Phys. Rev. Lett.}, 85:1887--1889, Aug 2000.

\bibitem{peters00}
M.~J. Peters, L.~E. McNeil, J.~Ping Lu, and D.~Kahn.
\newblock Structural phase transition in carbon nanotube bundles under
  pressure.
\newblock {\em Phys. Rev. B}, 61:5939--5944, 2000.

\bibitem{karmakar03}
Sukanta Karmakar, Surinder~M Sharma, P~V Teredesai, D~V~S Muthu, A~Govindaraj,
  S~K Sikka, and A~K Sood.
\newblock Structural changes in single-walled carbon nanotubes under
  non-hydrostatic pressures: x-ray and raman studies.
\newblock {\em New Journal of Physics}, 5(1):143, 2003.

\bibitem{merlenPRB05}
A.~Merlen, N.~Bendiab, P.~Toulemonde, A.~Aouizerat, A.~San Miguel, J.~L.
  Sauvajol, G.~Montagnac, H.~Cardon, and P.~Petit.
\newblock Resonant raman spectroscopy of single-wall carbon nanotubes under
  pressure.
\newblock {\em Phys. Rev. B}, 72(3):035409, 2005.

\bibitem{yao08}
Mingguang Yao, Zhigang Wang, Bingbing Liu, Yonggang Zou, Shidan Yu, Wang Lin,
  Yuanyuan Hou, Shoufu Pan, Mingxing Jin, Bo~Zou, Tian Cui, Guangtian Zou, and
  B.~Sundqvist.
\newblock Raman signature to identify the structural transition of single-wall
  carbon nanotubes under high pressure.
\newblock {\em Phys. Rev. B}, 78(20):205411, Nov 2008.

\bibitem{CharlierPRB96}
J.~C. Charlier, Ph. Lambin, and T.~W. Ebbesen.
\newblock Electronic properties of carbon nanotubes with polygonized cross
  sections.
\newblock {\em Phys. Rev. B}, 54(12):R8377--R8380, Sep 1996.

\bibitem{YildirimPRB00}
T.~Yildirim, O.~Gulseren, C.~Kilic, and S.~Ciraci.
\newblock Pressure-induced interlinking of carbon nanotubes.
\newblock {\em Phys. Rev. B}, 62:12648--12651, Nov 2000.

\bibitem{sluiterPRB02}
Marcel H.~F. Sluiter, Vijay Kumar, and Yoshiyuki Kawazoe.
\newblock Symmetry-driven phase transformations in single-wall carbon-nanotube
  bundles under hydrostatic pressure.
\newblock {\em Phys. Rev. B}, 65:161402, Apr 2002.

\bibitem{reichPRB02}
S.~Reich, C.~Thomsen, and P.~Ordej\'on.
\newblock Elastic properties of carbon nanotubes under hydrostatic pressure.
\newblock {\em Phys. Rev. B}, 65:153407, 2002.

\bibitem{CapazPSS04}
R.~B. Capaz, C.~D. Sparatu, P.~Tangney, M.~L. Cohen, and S.~G. Louie.
\newblock Hydrostatic pressure efects on the structural and eletronic
  properties of carbon nanotubes.
\newblock {\em phys. stat. sol.(b)}, 241:3352--3359, 2004.

\bibitem{tangney05}
P.~Tangney, R.~B. Capaz, C.~D. Spataru, M.~L. Cohen, and S.~G. Louie.
\newblock {\em Nano Letters}, 5(11), 2005.

\bibitem{ChoiPSS07}
I~H Choi, P~Y Yu, P~Tagney, and S~G Louie.
\newblock Vibrational properties of single walled carbon nanotubes under
  pressure from raman scattering experiments and molecular dynamics
  simulations.
\newblock {\em Physica Status Solidi}, 244(1):121--126, 2007.

\bibitem{arvanitidis05}
J.~Arvanitidis, D.~Christofilos, K.~Papagelis, K.~S. Andrikopoulos,
  T.~Takenobu, Y.~Iwasa, H.~Kataura, S.~Ves, and G.~A. Kourouklis.
\newblock Pressure screening in the interior of primary shells in double-wall
  carbon nanotubes.
\newblock {\em Phys. Rev. B}, 71(12):125404, 2005.

\bibitem{puechPRB06}
P~Puech, E~Flahaut, A~Sapelkin, H~Hubel, DJ~Dunstan, G~Landa, and WS~Bacsa.
\newblock Nanoscale pressure effects in individual double-wall carbon
  nanotubes.
\newblock {\em Phys. Rev. B}, 73(23), 2006.

\bibitem{puechPRB08}
P~Puech, A~Chandour, A~Sapelkin, C~Tinguely, E~Flahaut, D~J Dunstan, and
  W~Basca.
\newblock Raman g band in double-wall carbon nanotubes combining p doping and
  high pressure.
\newblock {\em Phys. Rev. B}, 78:045413, 2008.

\bibitem{kawasaki08}
S.~Kawasaki, Y.~Kanamori, Y.~Iwai, F.~Okino, H.~Touhara, H.~Muramatsu,
  T.~Hayashi, Y.~A. Kim, and M.~Endo.
\newblock Structural properties of pristine and fluorinated double-walled
  carbon nanotubes under high pressure.
\newblock {\em J. Phys. Chem. Solids}, 69(5-6):1203--1205, MAY-JUN 2008.
\newblock 14th International Symposium on Intercalation Compounds (ISIC 14),
  Seoul, SOUTH KOREA, JUN 12-15, 2007.

\bibitem{aguiar11}
A.~L. Aguiar, E.~B. Barros, R.~B. Capaz, A.~G.~Souza Filho, P.~T.~C. Freire,
  J.~Mendes Filho, D.~Machon, Ch. Caillier, Y.~A. Kim, H.~Muramatsu, M.~Endo,
  and A.~San-Miguel.
\newblock Pressure-induced collapse in double-walled carbon nanotubes: Chemical
  and mechanical screening effects.
\newblock {\em Journal of Physical Chemistry C}, 115:5378--5384, 2011.

\bibitem{ImtaniPRB07}
Ali~Nasir Imtani and V.~K. Jindal.
\newblock Structure of armchair single-wall carbon nanotubes under hydrostatic
  pressure.
\newblock {\em Phys. Rev. B}, 76(19):195447, Nov 2007.

\bibitem{ImtaniCMS08}
Ali~Nasir Imtani and V.~K. Jindal.
\newblock Bond lengths of armchair single-waled carbon nanotubes and their
  pressure dependence.
\newblock {\em Computational Materials Science}, 44:156--162, 2008.

\bibitem{ImtaniCMS09}
Ali~Nasir Imtani and V.~K. Jindal.
\newblock Structure of chiral single-walled carbon nanotubes under hydrostatic
  pressure.
\newblock {\em Computational Materials Science}, 46:297--302, 2009.

\bibitem{yangPRB07}
W.~Yang, R~Z Wang, X~M Song, B~Wang, and H~Yan.
\newblock Pressure-induced raman-active radial breathing mode transition in
  single-wall carbon nanotubes.
\newblock {\em Phys. Rev. B}, 75:045425, 2007.

\bibitem{SluiterPRB04}
Marcel H.~F. Sluiter and Yoshiyuki Kawazoe.
\newblock Phase diagram of single-wall carbon nanotube crystals under
  hydrostatic pressure.
\newblock {\em Phys. Rev. B}, 69(22):224111, Jun 2004.

\bibitem{christofilosDREL06}
D~Christofilos, J~Arvanitidis, C~Tzampazis, K~Papagelis, T~Takenobu, Y~Iwasa,
  H~Kataura, C~Lioutas, S~Ves, and G~A Kourouklis.
\newblock Raman study of metallic carbon nanotubes at elevated pressure.
\newblock {\em Diamond and Related Materials}, 15:1075--1079, 2006.

\bibitem{merlenPSS06}
A.~Merlen, P.~Toulemonde, N.~Bendiab, , A.~Aouizerat, J.~L. Sauvajol,
  G.~Montagnac, H.~Cardon, P.~Petit, and A.~San Miguel.
\newblock Raman spectroscopy of open-ended single wall carbon nanotubes under
  pressure: effect of the pressure transmiting medium.
\newblock {\em Phys. Stat. Sol.}, 243(3):690--699, 2006.

\bibitem{proctorPSS07}
J~E Proctor, M~P Halsall, A~Ghandour, and D~J Dunstan.
\newblock Raman spectroscopy of single-walled carbon nanotubes at
  high-pressures: Effect of interaction between the nanotubes and pressure
  transmitting media.
\newblock {\em Phys. Stat. Sol.}, 244(1):147--150, 2007.

\bibitem{AbouelsayedJPCC10}
A.~Abouelsayed, K~Thirunavukkuarasu, F~Hennrich, and C~A Kuntscher.
\newblock Role of the pressure transmitting medium for the pressure effects in
  single-walled carbon nanotubes.
\newblock {\em Journal of Physical Chemistry C}, 114:4424--4428, 2010.

\bibitem{elliot04}
James~A. Elliott, Jan~K. Sandler, Alan~H. Windle, Robert~J. Young, and Milo~S.
  Shaffer.
\newblock {\em Phys. Rev. Lett.}, 92, 2004.

\bibitem{YePRB2005}
X.~Ye, D.~Y. Sun, and X.~G. Gong.
\newblock Pressure-induced structural transition of double-walled carbon
  nanotubes.
\newblock {\em Phys. Rev. B}, 72:035454, Jul 2005.

\bibitem{yangAPL06}
X~Yang, G~Wu, and J~Dong.
\newblock Structural transformations of double-wall carbon nanotubes bundle
  under hydrostatic pressure.
\newblock {\em Applied Physics Letters}, 89(113101):113101--113103, 2006.

\bibitem{gadagkar06}
Vikram Gadagkar, Prabal~K. Maiti, Yves Lansac, A.~Jagota, and A.~K. Sood.
\newblock Collapse of double-walled carbon nanotube bundles under hydrostatic
  pressure.
\newblock {\em Phys. Rev. B}, 73(8):085402, Feb 2006.

\bibitem{brenner02}
Donald~W Brenner, Olga~A Shenderova, Judith~A Harrison, Steven~J Stuart, Boris
  Ni, and Susan~B Sinnott.
\newblock A second-generation reactive empirical bond order (rebo) potential
  energy expression for hydrocarbons.
\newblock {\em Journal of Physics: Condensed Matter}, 14(4):783, 2002.

\bibitem{brenner90}
D.~W. Brenner.
\newblock Empirical potential for hydrocarbons for use in simulating the
  chemical vapor deposition in diamond films.
\newblock {\em Phys. Rev. B}, 42(15), 1990.

\bibitem{Charlier96}
J.~C. Charlier, Ph. Lambin, and T.~W. Ebbesen.
\newblock Electronic properties of carbon nanotubes with polygonized cross
  sections.
\newblock {\em Phys. Rev. B}, 54(12):R8377--R8380, Sep 1996.

\bibitem{Liu05}
J.Z. Liu, Q.-S. Zheng, L.-F. Wang, and Q.~Jiang.
\newblock Mechanical properties of single-walled carbon nanotube bundles as
  bulk materials.
\newblock {\em Journal of the Mechanics and Physics of Solids}, 53(1):123 --
  142, 2005.

\bibitem{RuPRB00}
C.~Q. Ru.
\newblock Effective bending stiffness of carbon nanotubes.
\newblock {\em Phys. Rev. B}, 62(15):9973--9976, Oct 2000.

\bibitem{benedict98}
Lorin~X. Benedict, Nasreen~G. Chopra, Marvin~L. Cohen, A.~Zettl, Steven~G.
  Louie, and Vincent~H. Crespi.
\newblock Microscopic determination of the interlayer binding energy in
  graphite.
\newblock {\em Chemical Physics Letters}, 286:490--496, 1998.

\bibitem{sun04}
D.~Y. Sun, D.~J. Shu, M.~Ji~Feng Liu, M.~Wang, and X.~G. Gong.
\newblock {\em Phys. Rev. B}, 70, 2004.

\bibitem{wuPRB04}
Jian Wu, Ji~Zang, Brian Larade, Hong Guo, X.~G. Gong, and Feng Liu.
\newblock Computational design of carbon nanotube electromechanical pressure
  sensors.
\newblock {\em Phys. Rev. B}, 69:153406, Apr 2004.

\bibitem{caillier08}
Ch. Caillier, D.~Machon, A.~San-Miguel, R.~Arenal, G.~Montagnac, H.~Cardon,
  M.~Kalbac, M.~Zukalova, and L.~Kavan.
\newblock Probing high-pressure properties of single-wall carbon nanotubes
  through fullerene encapsulation.
\newblock {\em Physical Review B}, 77(12):125418, 2008.

\bibitem{sandler03}
J.~Sandler, M.~S.~P. Shaffer, A.~H. Windle, M.~P. Halsall, M.~A.
  Montes-Mor\'an, C.~A. Cooper, and R.~J. Young.
\newblock Variations in the raman peak shift as a function of hydrostatic
  pressure for various carbon nanostructures: A simple geometric effect.
\newblock {\em Phys. Rev. B}, 67:035417, Jan 2003.

\bibitem{freire07}
P.~T.~C. Freire, V.~Lemos, J.~A. Lima, G.~D. Saraiva, P.~S. Pizani, R.~O.
  Nascimento, N.~M. P.~S. Ricardo, J.~Mendes~Filho, and A.~G. Souza~Filho.
\newblock Pressure effects on surfactant solubilized single-wall carbon
  nanotubes.
\newblock {\em physica status solidi (b)}, 244(1):105--109, 2007.

\bibitem{Proctor06}
John~E. Proctor, Matthew~P. Halsall, Ahmad Ghandour, and David~J. Dunstan.
\newblock High pressure raman spectroscopy of single-walled carbon nanotubes:
  Effect of chemical environment on individual nanotubes and the nanotube
  bundle.
\newblock {\em Journal of Physics and Chemistry of Solids}, 67(12):2468--2472,
  2006.

\bibitem{PopovPRB02}
V.~N. Popov and Luc Henrard.
\newblock Breathinglike phonon modes of multiwalled carbon nanotubes.
\newblock {\em Phys. Rev. B}, 65:235415, May 2002.

\bibitem{christofilos07}
D~Christofilos, J.~Arvanitidis, G.~A. Kourouklis, S.~Ves, T.~Takenobu, Y~Iwasa,
  and H.~Kataura.
\newblock Identification of inner and outer shells of double-wall carbon
  nanotubes using high pressure raman spectroscopy.
\newblock {\em Phys. Rev. B}, 76, 2007.

\bibitem{puech04}
P.~Puech, H.~Hubel, D.~J. Dunstan, R.~R. Bacsa, C.~Laurent, and W.~S. Bacsa.
\newblock Discontinuous tangential stress in double wall carbon nanotubes.
\newblock {\em Phys. Rev. Lett.}, 93(9):095506, 2004.

\bibitem{PuechPSS04}
P~Puech, H~Hubel, D~J Dunstan, A~Bassil, R~Bacsa, A~Peigney, E~Flahaut,
  C~Laurent, and W~S Basca.
\newblock Light scattering of double wall carbon nanotubes under hydrostatic
  pressure: pressure effects on the internal and external tubes.
\newblock {\em phys. stat. sol.(b)}, 241:3360--3366, 2004.

\bibitem{Hanfland89}
M.~Hanfland, H.~Beister, and K.~Syassen.
\newblock Graphite under pressure: Equation of state and first-order raman
  modes.
\newblock {\em Phys. Rev. B}, 39:12598--12603, Jun 1989.

\bibitem{nicolle11}
Jimmy Nicolle, Denis Machon, Philippe Poncharal, Olivier Pierre-Louis, and
  Alfonso San-Miguel.
\newblock Pressure-mediated doping in graphene.
\newblock {\em Nano Letters}, 11(9):3564--3568, 2011.

\end{thebibliography}

\end{document}